\documentclass[12pt, a4paper]{article}
\usepackage{graphicx}
\usepackage{amssymb}
\usepackage{amsmath}
\usepackage{bm}
\usepackage{color}
\usepackage{cite}
\usepackage{slashed}
\usepackage{epstopdf}            
\usepackage{epsfig}
            
\newcommand{\beq}{\begin{equation}}
\newcommand{\eeq}{\end{equation}}
\newcommand{\ov}{\overline}

\newcommand{\ET}{\mbox{$\not \hspace{-0.10cm} E_T$ }}

\usepackage[colorlinks=true, linkcolor=black, citecolor=black,
urlcolor=black]{hyperref}

\begin{document}

\begin{titlepage}

\begin{flushright}

KIAS-16002

\end{flushright}

\vskip 2cm
\begin{center}

{\Large
{\bf 
Diphoton Excess at 750 GeV   in leptophobic 
\\ U(1)$^\prime$ model   inspired by $E_6$ GUT
}
}

\vskip 2cm

P. Ko$^{1}$,
Yuji Omura$^{2}$
and
Chaehyun Yu$^{3}$

\vskip 0.5cm

{\it $^1$School of Physics, KIAS, Seoul 02455, Kore
}\\[3pt]
{\it $^2$
Kobayashi-Maskawa Institute for the Origin of Particles and the
Universe, \\ Nagoya University, Nagoya 464-8602, Japan}\\[3pt]
{\it $^{3}$
Institute of Physics, Academia Sinica, Nangang, Taipei 11529, Taiwan}\\ [3pt]

\vskip 1.5cm

\begin{abstract}
We discuss the 750 GeV diphoton excess at the LHC@13TeV in the framework of
leptophobic U(1)$^\prime$ model inspired by $E_6$ grand unified theory (GUT).
In this model,  the Standard Model (SM)  chiral fermions carry charges under extra U(1)$^\prime$ gauge
symmetry which is spontaneously broken by a U(1)$^\prime$-charged singlet scalar  ($\Phi$).
In addition, extra quarks and leptons are introduced to achieve the anomaly-free conditions, which is a natural consequence of the assumed $E_6$  GUT.    These new fermions are vectorlike under the SM gauge group
but chiral under new U(1)$^\prime$,   and their masses come entirely from the nonzero vacuum expectation
value of $\Phi$ through the Yukawa interactions.
Then, the CP-even scalar $h_\Phi$ from $\Phi$ can be produced at the LHC by the gluon fusion and decay
to the diphoton via the one-loop diagram involving the extra quarks and leptons, and can be identified as  the origin of diphoton excess at 750 GeV.
In this model,  $h_\Phi$ can decay into a pair of dark matter particles as well as
a pair of scalar bosons, thereby a few tens of the decay width may be possible.  
\end{abstract}

\end{center}
\end{titlepage}

%%%%%%%%%%%%%%%%%%%%%%%%%%%%%%%%%%
\section{Introduction}
\label{sec:intro}
%%%%%%%%%%%%%%%%%%%%%%%%%%%%%%%%%%
The LHC Run-II experiment started taking data at the 13 TeV center mass of energy in 2015,
and searching new physics signals. One of the promising candidates for new physics  is 
an extra heavy scalar boson, which is predicted by many Beyond Standard Models (BSMs).
Therefore the search for such a heavy resonance 
plays a crucial role in testing BSMs.

Recently, both ATLAS and CMS collaborations reported excesses in the diphoton resonance 
search around 750 GeV~\cite{diphotonexcess}. 
The local (global) significances are $3.6 \, (2.0) \, \sigma$ (ATLAS) and  
$2.6 \, (1.2) \, \sigma$ (CMS), respectively. 
An interesting point is that the best fit value of the decay width of the resonance in the ATLAS data is 
$\sim$ 45 GeV, which is about 6\% of  the resonance mass, 
while the CMS data prefer a narrow decay width~\cite{diphotonexcess}. 
This issue is not still conclusive, but it may be important for theorists to survey the possibility of the 
diphoton resonance in BSMs~\cite{diphoton-composite,diphoton-dm,diphoton-vectorlike,Franceschini:2015kwy,diphoton-new,diphoton-axion,diphoton-stability,diphoton-susy,diphoton-u1,diphoton-ed,diphoton-2HDM,diphoton-higgcision}. 

One of the attractive candidates for the BSMs which predict the heavy scalar resonance is U(1)$^\prime$ extension of the Standard Model (SM). 
In a minimal setup, an additional gauged U(1)$^\prime$ is introduced and spontaneously broken by the nonzero vacuum expectation value (VEV) of an extra U(1)$^\prime$-charged scalar ($\Phi$).
In the effective lagrangian, an extra massive gauge boson is predicted and a CP-even scalar
also appears around the U(1)$^\prime$ breaking scale. If the scalar mixes with the SM Higgs, 
it will decay to diphoton, although the scalar mixing is strictly constrained by the
Higgs signal strengths and the other resonance searches at the collider experiments.
Besides, U(1)$^\prime$ may cause the gauge anomaly depending on the U(1)$^\prime$ 
charge assignments to the SM fermions.  Only anomaly-free and
generation-independent U(1)$^\prime$ charge 
assignment is the linear combination of the hyper-charge and  U(1)$_{\rm B-L}$, 
introducing the right-handed neutrinos.  However, such U(1)$^\prime$ faces a strong bound
from the Drell-Yan process, so that the U(1)$^\prime$ breaking scale can not be low.

As another possibility, we can discuss a U(1)$^\prime$ model, where the anomaly-free 
conditions are satisfied  by introducing extra chiral fermions. 
One interesting U(1)$^\prime$ extension would be the one motivated by 
the Grand Unified Theory (GUT) with a large rank. For instance, the $E_6$ GUT principally
predicts two additional U(1)$^\prime$ symmetries, together with extra chiral quarks and leptons.
The extra chiral fermions are vector-like under the SM gauge groups but chiral under the two 
U(1)$^\prime$.   Interestingly, it is suggested that a certain linear combination of the 
U(1)$^\prime$ charges could be leptophobic and, in fact, has been widely
studied so far in the many literatures
\cite{E6leptophobic,Babu,E6leptophobic2,Ko-0,Ko,Hisano,E6leptophobic3}.
Such a leptophobic U(1)$^\prime$ has been paid an attention to so far, because it can evade 
the strong bound from the Drell-Yan process. Then the U(1)$^\prime$ breaking scale can be lower 
than the U(1)$_{B-L}$ case.

In ref. \cite{Ko}, the present authors discussed the Higgs physics in the leptophobic U(1)$^\prime$ 
model  in the context of Type-II 2HDM with extra leptophobic U(1)$^\prime$ instead of softly broken 
$Z_2$ symmetry \cite{Ko-0}.   The reason for 2HDM is that the SM up-type and the down-type fermions 
carry different U(1)$^\prime$ charges,  so that one can not write Yukawa couplings for all the SM fermions 
with one Higgs doublet. 
It is mandatory to extend the Higgs sector by introducing (at least) one more Higgs doublet where  
two Higgs doublets $H_1$  and $H_2$ carry different U(1)$^\prime$ charges.
This way one can write down all the Yukawa couplings for the SM fermions without generating dangerously 
huge FCNC mediated by neutral Higgs boson \cite{Ko-0}.
They also found a neutral fermion ($\psi_X$) in the extra fermions make a good dark matter candidate 
in this model \cite{Ko}.
One generation of the SM fermions and extra fermions belongs to a single ${\bf 27}$ representation 
of $E_6$  gauge group, making an anomaly free representation. 

In this model, both dark matter and extra chiral fermions get their masses entirely from the nonzero 
VEV of $\Phi$  and the Yukawa couplings.
One interesting aspect is that the extra quarks only couple with $\Phi$, but not Higgs doublets because of 
the  U(1)$^\prime$ charge assignment.
Then,  the CP-even scalar mode ($h_\Phi$), which appears after the U(1)$^\prime$ symmetry breaking,
does not couple with the SM fermions, neglecting mixing between $h_\Phi$
and the other scalars from two Higgs doublets.

Once we assume that the mixing is tiny, we can expect that  $h_\Phi$ mainly decays through the Yukawa 
couplings with the extra chiral fermions.
As a result, $h_\Phi$ can decay to $gg$ and $\gamma \gamma$ via the one-loop diagrams
involving  the extra chiral fermions which are vector-like under the SM gauge group.   In addition, $h_\Phi$
can decay into a pair of DM as well as a pair of scalar bosons ($h h, H h, HH, AA$) 
if kinematically allowed, thereby its decay width being increased significantly. 
This is how we explain the 750 GeV  diphoton excess reported by ATLAS and CMS recently.
We investigate the parameter region favored by the 750 GeV diphoton excess, and discuss
the consistency with the other results on the new physics search at the LHC.

In section~\ref{sec;setup}, we introduce our model setup inspired by $E_6$ GUT.
In section~\ref{sec;signal}, we perform phenomenological analysis on the 750 GeV diphoton 
excess in our model, and discuss the dark matter physics, according to the interpretation 
of the diphoton  excess.  Finally we summarize our results and give  some future prospects 
for probing the  scenarios in the future LHC experiments in section~\ref{sec:summary}.

%%%%%%%%%%%%%%%%%%%%%%%%%%%%%%%%%%%%%%%%%%%%%%%%%%%%%%%
\section{Leptophobic U(1)$^\prime$ model inspired by $E_6$ GUT}
\label{sec;setup}
%%%%%%%%%%%%%%%%%%%%%%%%%%%%%%%%%%%%%%%%%%%%%%%%%%%%%%%%

\subsection{Model: matter contents and their quantum numbers}
\label{matter}
Here, we introduce our setup based on refs. \cite{Ko-0,Ko}, where the SM gauge group is extended to include 
leptophobic  U(1)$^\prime$ gauge symmetry, and additional chiral fermions are introduced in order to cancel 
gauge anomalies.   
From the bottom-up point of view, there are many varieties for U(1)$^\prime$ charge assignments 
that achieve gauge anomaly cancellations.
One simple and interesting way is to consider U(1)$^\prime$ symmetries predicted by the supersymmetric 
$E_6$ GUT,   where the all SM fields including Higgs doublets can be unified into three-family 
${\bf 27}$-dimensional fields.

Our approach is bottom-up, but let us also explain the underlying theory briefly.
The $E_6$ GUT predicts additional two U(1) symmetries in addition to the SM gauge groups.
After the $E_6$ breaking, one ${\bf 27}$-dimensional fermion field can be decomposed to all the SM 
fermions in one generation,   together with right-handed and left-handed neutrinos, and one extra 
SM vector-like quark and lepton pairs.   
If U(1)$^\prime$, which is given by a linear combination of the two U(1), 
is assumed to be remained up to the low scale, the anomaly-free condition for the extra U(1)$^\prime$
can be satisfied by the chiral fermions predicted by the ${\bf 27}$-dimensional fields.

%In the framework of the supersymmetric $E_6$ GUT, which we expect to be our underlying theory, 
%there are also bosonic partners of the fermions. 
%For instance, the SM Higgs doublet could correspond to the bosonic partner of the extra lepton.
%Besides, the bosonic partner of the neutrino is charged under U(1)$^\prime$, but not the SM gauge groups. 
%Then, it works as the field, $\Phi$, to break U(1)$^\prime$. 
%Our approach is the bottom-up one, so we simply discuss the diphoton excess and phenomenology concerned with the 
%excess, using the fermions and some bosonic fields which are predicted by the ${\bf 27}$-dimensional chiral superfields
%in the supersymmetric $E_6$ GUT. 
%We do not touch the detail of how to realize our U(1)$^\prime$ model anymore, but we give some comments on the connection between our phenomenological model
%and the supersymmetric $E_6$ GUT later. Let us introduce our setup of $U(1)^\prime$ non-supersymmetric model below.

In our model inspired by the $E_6$ GUT, extra U(1)$^\prime$ is introduced and its charges are 
assigned to the SM particles  by the linear combination of U(1)$_Y$ and two U(1) of $E_6$, called 
U(1)$_\chi$ and U(1)$_\psi$. 
It is known that one certain linear combination realizes a so-called leptophobic U(1)$^\prime$,
where the SM charged leptons which are not charged under the U(1)$^\prime$\cite{E6leptophobic,Babu,E6leptophobic2,Ko-0,Ko,Hisano,E6leptophobic3}.   
We adopt this leptophobic charge assignments and discuss the phenomenology.
The U(1)$^\prime$ charge assignment for the SM fermions is shown in table~\ref{table1}. 
The U(1)$^\prime$ is the subgroup of $E_6$, so that it is anomaly-free introducing the 
extra chiral fermions predicted by the ${\bf 27}$-dimensional fields (see for example  \cite{E6review2}).

The quantum numbers of the SM and the extra chiral fermions are summarized in table~\ref{table1}. 
$Q^i$, $d_R^{ i}$ and  $u_R^{i}$   are the SM quarks, which carry nonzero U(1)$^\prime$ charges, 
while $L^i$ and $e_R^{ i}$ are the SM leptons which are not charged under the U(1)$^\prime$, as shown in 
table~\ref{table1}.    In addition, extra quarks and leptons are contained in the ${\bf  27}$, which are denoted 
by $D^i_L$, $D^{ i}_R$ and  $\widetilde{H}^i_L$, $\widetilde H^{i}_R$ $(i=1, \,2, \, 3)$. 
There are left-handed and right-handed neutrinos, 
$n^{ i}_R$, $N^i_L$, whose U(1)$^\prime$ charges are $ \pm 1$.
Note that all the matter fields (both fermions and bosons) carry nonzero gauge charges and there are 
no gauge  singlets in this model.
%\footnote{ Note that all of the fields are embedded into a ${\bf 27}$-representational 
%superfield for the each generation in the supersymmetric extension of this model.}.

%%%%%%%%%%%%%%%%%%%%%%%%%%%%%%%%%%%%%%%%%%%%%%%%%%%%%%%%%%%
\begin{table}[t]
\caption{Matter contents in U(1)$^\prime$ model inspired by E$_6$ GUTs. Here, $i$
 denotes the generation index: $i=1, \, 2, \, 3$. }
 \label{table1}
\begin{center}
\begin{tabular}{ccccccc}
\hline
\hline
Fields& spin    & ~~$\text{SU}(3)$~~ & ~~$\text{SU}(2)$~~  & ~~$\text{U}(1)_Y$~~ & ~~$\text{U}(1)'$~~ & ~~$Z^{\rm ex}_2$~~     \\ \hline  
 ~~$Q^i$~~& & ${\bf 3}$         &${\bf 2}$      &         $1/6$              &    $-1/3$ &      \\
  $u^{i}_{R}$&  & ${\bf 3}$         &${\bf1}$      &       $2/3$        &  $2/3$ &         \\  
   $d^{ i}_R$&  &  ${\bf 3}$        &${\bf1}$      &         $-1/3$              &    $-1/3$ &        \\  
   $L_i$ & 1/2   & ${\bf1}$         &${\bf 2}$      &            $-1/2$        &      $0$  & + \\ 
  $e^{ i}_R$ &  & ${\bf1}$         &${\bf1}$      &           $-1$            &   $0$ &     \\ 
        $n^{ i}_R$ &  & ${\bf1}$         &${\bf1}$      &           $0$            &    $1$&      \\  \hline 
    $H_2$ & & ${\bf1}$         &${\bf 2}$      &          $-1/2$            &   $0$  &         \\   
   $H_1$ & 0 & ${\bf1}$         &${\bf 2}$      &          $-1/2$            &   $-1$   &    +   \\  
      $\Phi$ & & ${\bf1}$         &${\bf1}$      &           $0$            &    $-1$ &   \\ \hline
     $D^i_{L}$ &  & ${\bf 3}$         &${\bf1}$      &       $-1/3$        &  $2/3$  &      \\ 
    $D^{ i}_{R}$ &  & ${\bf 3}$         &${\bf1}$      &       $-1/3$        &  $-1/3$ &          \\ 
   $\widetilde H^{i}_L$ & 1/2  & ${\bf1}$         &${\bf 2}$      & $-1/2$        &      $0$ & $-$    \\ 
  $\widetilde H^{i}_R$ & & ${\bf1}$         &${\bf 2}$      &           $-1/2$            &   $-1$ &     \\  
   $N^i_L$ &  & ${\bf1}$         &${\bf1}$      &           $0$            &    $-1$  &    \\  \hline \hline
\end{tabular}
\end{center}
\end{table}

In order to realize electroweak symmetry breaking and U(1)$^\prime$ breaking,
we introduce scalars: two Higgs doublets, denoted by $H_1$ and $H_2$, and a U(1)$^\prime$-charged
singlet, $\Phi$. Their quantum numbers are shown in table~\ref{table1}.  
Note that scalar fields have nothing to do with $E_6$, since they do not belong to ${\bf 27}$-representation 
of $E_6$.  We have used the $E_6$ connection only to the fermion sector in relation with gauge anomaly
cancellation. Scalar fields do not contribute to gauge anomaly, and thus need not belong to ${\bf 27}$ of 
$E_6$ within the bottom-up approach we are taking in this work.
As we discuss in the subsection \ref{Yukawa}, $H_1$ and $H_2$ are distinguished 
by their U(1)$^\prime$ charges and couple with the up-type and the down-type fermions, respectively.
$\Phi$ develops the nonzero VEV, and breaks U(1)$^\prime$ gauge symmetry spontaneously. 
Note that our neutrino is a Dirac particle and the mass term is given by the Yukawa coupling: 
$y^{ij}_\nu \overline{L_i} H_1 n_{R \, j}$. 

Note that our model is nonsupersymmetric by construction.
Here let us give brief comments on the connection between our nonsupersymmetric U(1)$^\prime$ model
and the supersymmetric $E_6$ GUT, which we might expect to be the underlying theory of our model. 
In the framework of the supersymmetric $E_6$ GUT,  one ${\bf 27}$-representational chiral superfield 
realizes all the fermions in one generation shown in table \ref{table1}.    In addition, the bosonic partners
with opposite $R$-parity are predicted. As we can see in table \ref{table1}, 
the quantum numbers of $H_1$ and $H_2$ are the same as the ones of $\widetilde H_L$ and 
$\widetilde H_R$  and the U(1)$^\prime$ charge of $\Phi$ is the same as $ \overline{n^i_R}$ and $N^i_L$.
In fact, the U(1)$_\chi$ and U(1)$_\psi$ charges of $\Phi$, which are introduced in ref. \cite{Ko}, 
are the same as the ones of $N^i_L$.   Then, the bosonic fields could be interpreted as the superpartners 
of the extra fermions, although we have to consider an issue how to decouple the other 
bosonic fields  except  $H_{1,2}$ and $\Phi$, which is known to be notoriously difficult in any GUT models.

Another issue may be how to 
realize the U(1)$^\prime$ in the low energy regime, although our approach is bottom-up. 
The possibility that the  leptophobic U(1)$^\prime$ is generated by kinetic mixing in the 
supersymmetric $E_6$ GUT has been studied in  ref.~\cite{Babu}. 
Furthermore, Yukawa couplings may cause serious  problems in not only E$_6$ but also SO(10) and 
SU(5) GUTs. The GUTs  unify the matter fields in the elegant ways, but the unification makes
it harder to explain the realistic fermion mass matrices. 

In this work, our main motivation is to propose one possible setup to 
explain the diphoton excess, so that we take the bottom-up approach.
Our matter contents and U(1)$^\prime$ charge assignment
are inspired by the supersymmetric $E_6$ GUT, but the study about the
explicit consistency with the GUT is clearly beyond our scope.
In a sense, our model setup could be considered as a simplified model for $E_6$ GUT, 
where fermions form a complete ${\bf 27}$-representation of $E_6$, but the scalar fields do not.
We introduce only a minimal number of scalar fields in order to break the gauge symmetry spontaneously
and generate masses for the SM and extra chiral fermions in our model, in such a way that the neutral 
Higgs-mediated FCNC is not present at tree level \cite{Ko-0}.

Note that we also introduce a new discrete symmetry $Z_2^{\rm ex}$ in table \ref{table1}, and assign 
positive parity  to the SM fermions and the negative parity to extra fermions. $\Phi$ and $H_{1,2}$ are 
$Z_2^{\rm ex}$-even.    As discussed in ref. \cite{Ko},  the $Z^{\rm ex}_2$ symmetry is defined by the 
subgroup of U(1)$_\chi$ or U(1)$_\psi$ multiplied by $(-1)^{2s}$,   where $s$ is the spin of the matter fields.\footnote{This corresponds to the R-parity in the minimal supersymmetric SM.}
That is, it is anomaly-free and can be realized by one proper Higgs field to break the GUT symmetry. 
From the phenomenological point of view, this $Z_2^{\rm ex}$ symmetry is required to avoid the tree-level 
FCNCs,  forbidding the tree-level mass mixing between the SM quarks and the extra quarks.

Below, we explain the relevant couplings in our U(1)$^\prime$ model, 
where there are only two Higgs doublets and $\Phi$, together
with the chiral fermions for the anomaly-free conditions, as shown in table \ref{table1}.

\subsection{Extra leptophobic gauge boson}
We consider a U(1)$^\prime$ model with two Higgs doublets, where only $H_1$ carries
nonzero U(1)$^\prime$ charge.
The kinetic term of $H_1$ has an extra term associated with the U(1)$^\prime$ gauge
boson $\hat{Z}_H$,
%-----------------
\begin{equation}
D^\mu H_1 = D_\mu^\textrm{SM} H_1 - i g_H \hat{Z}_{H}^\mu H_1
\end{equation}
%-----------------
with being the U(1)$^\prime$ gauge coupling $g_H$, while that of $H_2$ has
only the SM part. The mass matrix of $\hat{Z}$ and $\hat{Z}_H$ is given by
%-----------------
\begin{equation}
{\cal M}_{\hat{Z},\hat{Z}_H}^2 = 
\left(
\begin{array}{cc}
g_Z^2 v_H^2 & - g_Z g_H v_1^2 \\
- g_Z g_H v_1^2 & g_H^2 (v_1^2 + v_\Phi^2)
\end{array}
\right),
\end{equation}
%-----------------
where the mixing angle $\xi$ between $\hat{Z}$ and $\hat{Z}_H$ is given by
%-----------------
\begin{equation}
\tan 2 \xi = -\frac{2 g_Z g_H v_1^2}{g_H^2(v_1^2+v_\Phi^2) - g_Z^2 v_H^2}.
\end{equation}
%-----------------
$g_Z$ is the gauge coupling in the SM: $g_Z=\sqrt{g'^2+g^2}$, where $g'$ and $g$ are
the $U(1)_Y$ and $SU(2)_L$ gauge couplings. 
The new gauge boson $Z_H$ is constrained by the collider searches and the electroweak precision
tests.  In ref.~\cite{Ko}, the present authors studied them, finding out that $g_H \lesssim 0.1$ for
$M_{Z_H} \gtrsim 400$ GeV, and $g_H \approx 0.01$ for $M_{Z_H} \sim 200$ GeV 
(see section IV.C.1 and figure~\ref{fig1} in ref.~\cite{Ko} for more detail).  For such a small $g_H$, the $Z_H Z_H $ fusion into $h_\Phi$ 
would be small, and shall be ignored in this paper.

\subsection{Scalar sector}
In our model, there are two Higgs doublets, $H_1$ and $H_2$, 
where $H_1$ ($H_2$) gives masses to up-type quarks (down-type quarks and charged leptons) after 
EW symmetry breaking, and one singlet scalar $\Phi$ with nonzero U(1)$^\prime$ charge.
$H_1$ is charged under the extra U(1)$^\prime$ symmetry, which could be the origin of the $Z_2$ symmetry 
of the usual 2HDM,   while $H_2$ is uncharged. Then, the so-called $\mu$ term of the Higgs potential,
$\mu H_1^\dagger H_2$ which breaks $Z_2$ softly, is not invariant under the U(1)$^\prime$ symmetry.
This $\mu$-term is replaced by  $\mu_\Phi H_1^\dagger H_2 \Phi$, and the $\mu$ term will be 
generated after U(1)$^\prime$ symmetry breaking. Note that 2HDMs with softly broken $Z_2$ symmetry 
are discussed widely. 
This U(1)$^\prime$, which distinguishes two Higgs doublets, suggests
the origin of  $Z_2$ symmetry~\cite{Ko-0}
(see ref.~\cite{Huang:2015wts} for implementation to $SU(2)_H$ gauge symmetry). 

The potential of the scalar fields in our model  is given by
%-----------------
\begin{eqnarray}
V_{\rm scalar}&=& \tilde{m}_1^2 H_1^\dagger H_1
+\tilde{m}_2^2 H_2^\dagger H_2
+ \frac{\lambda_1}{2} \left( H_1^\dagger H_1 \right)^2
+ \frac{\lambda_2}{2} \left( H_2^\dagger H_2 \right)^2
\nonumber \\
&&
+ \lambda_3 H_1^\dagger H_1 H_2^\dagger H_2
+ \lambda_4 H_1^\dagger H_2 H_2^\dagger H_1
+V_\Phi,
\end{eqnarray}
%-----------------
where the potential containing the singlet scalar $\Phi$ is
%-----------------
\begin{eqnarray}
V_\Phi &=&
\tilde{m}_\Phi^2 \Phi^\dagger \Phi
+\frac{\lambda_\Phi}{2} \left( \Phi^\dagger \Phi \right)^2
- \left( \mu_\Phi H_1^\dagger H_2 \Phi + \textrm{h.c.} \right)
+ \tilde{\lambda}_1 H_1^\dagger H_1 \Phi^\dagger \Phi
+ \tilde{\lambda}_2 H_2^\dagger H_2 \Phi^\dagger \Phi.
\end{eqnarray}
%-----------------
After spontaneous symmetry breaking, the scalar fields can be expanded
around their vacuum expectation values as
%-----------------
\begin{equation}
H_i = \left(
\begin{array}{c}
\phi_i^+ \\
\frac{1}{\sqrt{2}}(v_i+h_i+i \chi_i)
\end{array}
\right),
~
\Phi = \frac{1}{\sqrt{2}}(v_\Phi + h_\Phi + i \chi_\Phi),
\end{equation}
%-----------------
where $v_1=v_H \cos \beta$, $v_2=v_H \sin \beta$
and $v_H = 246$ GeV. The neutral CP-even scalars
generally mix with each other. The mass matrix is given by
%-----------------
\begin{equation}
\tilde{\cal M}^2 =
\left(
\begin{array}{ccc}
\tilde{\cal M}_{11}^2 & \tilde{\cal M}_{12}^2 & \tilde{\cal M}_{1 \Phi}^2 \\
\tilde{\cal M}_{12}^2 & \tilde{\cal M}_{22}^2 & \tilde{\cal M}_{2 \Phi}^2 \\
\tilde{\cal M}_{1 \Phi}^2 & \tilde{\cal M}_{2 \Phi}^2 & \tilde{\cal M}_{ \Phi \Phi}^2
\end{array}
\right)
\end{equation}
%-----------------
where
%-----------------
\begin{eqnarray}
\tilde{\cal M}_{11}^2 &=&
\frac{1}{2}\lambda_1 v_H^2 \cos^2\beta
+\frac{1}{2\sqrt{2}} \mu_\Phi v_\Phi \tan\beta,
\\
\tilde{\cal M}_{22}^2 &=&
\frac{1}{2}\lambda_2 v_H^2 \sin^2\beta
+\frac{1}{2\sqrt{2}} \mu_\Phi v_\Phi \cot\beta,
\\
\tilde{\cal M}_{\Phi\Phi}^2 &=&
\frac{1}{2}\lambda_\Phi v_\Phi^2
+\frac{1}{2\sqrt{2}} \frac{\mu_\Phi v_H^2}{v_\Phi} \sin\beta\cos\beta,
\\
\tilde{\cal M}_{12}^2 &=&
-\frac{1}{\sqrt{2}}\mu_\Phi v_\Phi
+(\lambda_3+\lambda_4) v_H^2 \sin\beta \cos\beta,
\\
\tilde{\cal M}_{1\Phi}^2 &=&
\tilde{\lambda}_1 v_H v_\Phi \cos\beta 
- \frac{1}{\sqrt{2}} \mu_\Phi v_H \sin\beta,
\\
\tilde{\cal M}_{2\Phi}^2 &=&
\tilde{\lambda}_2 v_H v_\Phi \sin\beta 
- \frac{1}{\sqrt{2}} \mu_\Phi v_H \cos\beta.
\end{eqnarray}
%-----------------
Since the recent data at the LHC imply that the 125 GeV scalar boson
is almost the SM-like Higgs boson, the mixing between
$h_{1,2}$ and $h_\Phi$ must be small. For simplicity we assume that
there is no mixing between them by setting
%-----------------
\begin{eqnarray}
\tilde{\lambda}_1 =  \frac{\mu_\Phi}{\sqrt{2}v_\Phi} \tan\beta, ~
\tilde{\lambda}_2 =  \frac{\mu_\Phi}{\sqrt{2}v_\Phi} \cot\beta.
\end{eqnarray}
%-----------------
The mass of $h_\Phi$ is determined by $m_{h_\Phi}^2 = \tilde{\cal M}_{\Phi\Phi}^2$.

The other two CP-even Higgs bosons, $h_1$ and $h_2$ mix with each other and
we identify the light boson as the SM-like Higgs boson $h$ with the mass
$m_h = 125$ GeV while the other one is the heavy Higgs boson $H$:
%-----------------
\begin{equation}
\left(
\begin{array}{c}
h_1 \\ h_2
\end{array}
\right)
=
\left(
\begin{array}{cc}
\cos \alpha_h & -\sin \alpha_h \\
\sin \alpha_h & \cos \alpha_h
\end{array}
\right)
\left(
\begin{array}{c}
H \\ h
\end{array}
\right),
\end{equation}
%-----------------
with the mixing angle $\alpha_h$.
Since the properties of the Higgs boson measured at the LHC is consistent with
the SM prediction, we assume that 
the decoupling limit, $\sin(\beta-\alpha_h)=1$,
is realized in the two Higgs doublet sector.
Note that we still have to keep
two Higgs doublets in order to write down the Yukawa couplings for all the observed SM chiral fermions.

\subsection{Yukawa sector of extra fermions}
\label{Yukawa}
Finally, we introduce our Yukawa couplings involving the scalars and the fermions in table \ref{table1}.
The U(1)$^\prime$-symmetric Yukawa couplings in our model  
are given by  
\begin{equation}
\label{potential}%
V_y= y^u_{ij}\overline{u^{ j}_R} H^{\dagger}_1 i \sigma_2  Q^i  + y^d_{ij} \overline{d^{ j}_R} H_2 Q^i + y^e_{ij}\overline{e^{ j}_R}H_2 L^i + y^n_{ij} \overline{n^{ j}_R} H^{\dagger}_1 i \sigma_2 L^i  
+  H.c.,
\end{equation}
where $\sigma_2$ is the Pauli matrix.
The Yukawa couplings to generate the mass terms for the extra particles
are 
\begin{equation}
\label{potential2}
V^{\rm ex}= y^D_{ij} \overline{D_R^{j}} \Phi  D_L^i  +y^H_{ij}   \ov{ \widetilde H_R^{ j}} \Phi  \widetilde H_L^i + y^N_{IJ} \overline{N^c_L}  H_1^\dagger i \sigma_2 \widetilde{H}^i_L  
+y'^N_{IJ}   \overline{\widetilde{H}_R^{ i}}  H_2 N^j_L +  H.c. \ .
\end{equation}
Let us comment on the mass spectrum derived from $V_y$ and $V^{\rm ex}$.
$H_1$, $H_2$ and $\Phi$ develop nonzero VEVs, and break $\text{SU}(2)_L\otimes \text{U}(1)_Y$
and U(1)$^\prime$ symmetries.
The extra colored and charged particles obtain heavy masses from the nonzero VEV of $\Phi$. 
We also find the neutral particle  masses are generated by the VEVs of Higgs doublets and $\Phi$. 
These massive extra particles are $Z^{\rm ex}_2$-odd, and thus the lightest neutral fermion among them 
becomes stable and could be a good cold dark matter candidate \cite{Ko}.  
The detailed phenomenological study of the fermionic DM $\psi_X$(which is mostly $n_L$) scenario is 
presented in ref.~\cite{Ko}.   
As we mentioned in subsection \ref{matter} and ref. \cite{Ko}, 
$Z^{\rm ex}_2$ is the subgroup of $E_6$, which is either U(1)$_\psi$ or U(1)$_\chi$, and
might also be generated by the $E_6$ gauge group.
  
On the other hand, the charged extra leptons  decay to the extra neutral particles and charged leptons, 
and the  colored extra quarks decay to the extra neutral ones and the SM particles through the 
higher-dimensional operators \cite{Ko}.   The direct search for the extra particles at the LHC imposes 
the lower  bounds on their masses. Their signals are colored or charged particles with large missing energy, 
so that the current lower mass bounds are  about $400-800$~GeV \cite{Bound-ExtraQuark,
Bound-ExtraLepton,Bound-ExtraLepton2}.  However  we have to keep in mind that 
these bounds depend on the dark matter mass, and thus are quite model dependent. 

\subsection{Scalar DM \label{scalarDM}}
Before moving to our phenomenology, let us suggest another possibility of the dark matter, adding an extra scalar, $X$.
Without any theoretical problem, one can introduce a new $Z_2^{\rm ex}$-odd scalar field $X$ with the $SU(3)_C \times SU(2)_L \times 
U(1)_Y \times U(1)'$ quantum numbers equal to  $(1,1,0;-1)$. 
The quantum number of $X$ is the same as $\overline{n^i_{R}}$,
and the $Z_2^{\rm ex}$ charge of $X$ is given by the $Z_2^{\rm ex}$ charge of the neutrino multiplied by $(-1)^{2s}$.
That is, $X$ could be interpreted as the superpartner of $\overline{n^i_{R}}$, in the framework of
the supersymmetric $E_6$ GUT. 

%%%%%%%%%%%%%%%%%%%%%

The gauge-invariant Lagrangian 
involving  $X$ is given by 
\begin{eqnarray}
{\cal L}_{X} & = & D_\mu X^\dagger D^\mu X - 
( m_{X0}^2 + \lambda_{H_1 X} H_1^\dagger H_1 + \lambda_{H_2 X} H_2^\dagger H_2  ) 
X^\dagger X - \lambda_{X} ( X^\dagger X )^2 
\nonumber \\
& - & \left( \lambda_{\Phi X}^{''} ( \Phi^\dagger X )^2 + H.c.
\right) - \lambda_{\Phi X} \Phi^\dagger \Phi X^\dagger X 
- \lambda_{\Phi X}^{'} | \Phi^\dagger X |^2
\nonumber  \\
& - & \left( y_{dX}^D  \overline{d_R} D_L X + y_{LX}^{\tilde{H}}
\overline{L} \widetilde{H}_R X^\dagger + H.c. \right).
\end{eqnarray}
Generation indices are suppressed for simplicity, but should be included in actual calculation.
We have imposed $Z_2^{\rm ex}$ symmetry, which forbids dangerous terms such as  
\[
\Phi^\dagger X, \, \, \,  H_1^\dagger H_1 \Phi^\dagger X , \, \, \, etc.
\]
that would make $X$ decay.
Assuming that $\langle X \rangle = 0$, $X$ would be stable and make another good 
candidate of CDM, in addition to a neutral fermion DM discussed in the previous subsection.
Since $X$ feels U(1)$^\prime$ gauge force, it is a baryonic DM and interact with the nuclei 
through $Z_H$ exchanges.

The exotic quark $D_L$ will decay into $\bar{d}_R + \bar{X}$ through the
$y_{dX}^D$ term.   The collider signature will be dijet + missing $E_T$ and is  similar to
the squark search bounds.  Likewise, the exotic leptons $\tilde{H}_R$ can decay into
$l + X$.  The collider signature will be dilepton + missing $E_T$ and is  similar to
the slepton search bounds.  Note that the fermionic DM $\psi_X$ (mostly composed of $n_L$)  
discussed in the previous subsection can  decay into $X + \nu$ if it is kinematically allowed.
Therefore the lighter one of $X$ and $\psi_X \approx n_L$ would be a good DM candidate. 

%%%%%%%%%%%%%%%%%%%%%%%%%%%%%%%%%%%%%%%%%%%%%%%%%%%%%%%
\section{Phenomenology}
\label{sec;signal}
%%%%%%%%%%%%%%%%%%%%%%%%%%%%%%%%%%%%%%%%%%%%%%%%%%%%%%%%
As we have mentioned, we assume that the diphoton excess around 750 GeV is
interpreted as the resonant production of the CP-even scalar $h_\Phi$.
In the simple setup in section~\ref{sec;setup}, $h_\Phi$ does not interact
with the SM fermions at the tree level since there is no mixing with $h_i$.
The ATLAS data prefer rather a large decay width ($\Gamma \sim 45$ GeV) of $h_\Phi$ , but
we note that the narrow width of $h_\Phi$ where $\Gamma_\Phi \lesssim 1\% \times m_\Phi$ is not 
excluded either by the data.   In fact the CMS data favor a narrower width, in contrast to the ATLAS data.
This issue is very important for the model buildings when the 750 GeV diphoton excess is confirmed 
in the future.   

In the following analysis, we will simply assume that $\Gamma_\textrm{tot} = 10$ GeV. 
We note that the decay width for $h_\Phi \to gg$, $\gamma\gamma$, or any SM particles is much less 
than $O(10)$ GeV.   In order to enhance the decay width, the exotic decay channels of $h_\Phi$ will be 
assumed. One of the candidate is the $h_\Phi$ decay into a pair of  fermion DM, which exists in the model 
naturally.  The decay into a pair of the scalar DM, which is introduced 
in section~\ref{scalarDM}, could be another candidate.

When the gluon fusion  is dominant for the $h_\Phi$ production,
the cross section for the diphoton production via the $h_\Phi$ resonance
can be described in terms of the decay widths of $h_\Phi \to gg$
and $h_\Phi \to \gamma\gamma$ and 
the integral of parton distribution functions (pdfs) of gluons ($C_{gg}$) 
by
%-----------------
\begin{equation}
\sigma(gg\to h_\Phi \to \gamma\gamma)=
\frac{C_{gg}}{s m_{h_\Phi} \Gamma_\textrm{tot}} \Gamma[h_\Phi\to gg]
\Gamma[h_\Phi \to \gamma\gamma],
\end{equation}
%-----------------
where 
%-----------------
\begin{equation}
C_{gg}=\frac{\pi^2}{8}\int_{\tau}^{1}
\frac{d x}{x}
g\left( x, m_\Phi^2 \right)
g\left(\frac{\tau}{x}, m_\Phi^2\right),
\end{equation}
%-----------------
with $\tau=m_\Phi^2/s$ and $g(x,Q^2)$ is the gluon pdf at $x=Q^2$.
Numerically, 
$C_{gg}=2137$ for LHC@13TeV and $C_{gg}=174$ for LHC@8TeV~\cite{Franceschini:2015kwy}.

The decay rates of $h_\Phi$ to two gluons and two photons are given by
%-----------------
\begin{eqnarray}
\Gamma[h_\Phi \to gg] &=&
\frac{\alpha_s^2 m_{h_\Phi}^3}{128 \pi^3 v_\Phi^2}
\left|
\sum_{q^\prime}^{} A_{1/2}^H (\tau_{q^\prime})
\right|^2,
\\
\Gamma[h_\Phi \to \gamma\gamma] &=&
\frac{\alpha^2 m_{h_\Phi}^3}{256 \pi^3 v_\Phi^2}
\left| \sum_{q^\prime}^{} N_c Q_{q^\prime}^2 A_{1/2}^H (\tau_{q^\prime})
+\sum_{l^\prime}^{} Q_{l^\prime}^2 A_{1/2}^H (\tau_{l^\prime})
+\frac{v_H v_\Phi}{2 m_{H^\pm}^2} 
\lambda_{h_\Phi H^+ H^-} A_0^H (\tau_{H^\pm})
\right|^2, 
\nonumber \\
&&
\end{eqnarray}
%-----------------
where 
$\tau_i = m_{h_\Phi}^2/4 m_i^2$ and $q^\prime (l^\prime)$ are the extra
quarks (charged leptons), respectively.
The loop functions are defined by
%-----------------
\begin{eqnarray}
A_{1/2}^H (\tau) &=& 2 [ \tau + (\tau - 1) f(\tau) ] / \tau,
\\
A_0^H(\tau) &=& -[\tau - f(\tau)]/\tau^2 , 
\end{eqnarray}
%-----------------
where the fucntion $f(x)$ is defined by
%-----------------
\begin{equation}
f(x) = \left\{ 
\begin{array}{ccc}
\arcsin^2 \sqrt{x} &,& \textrm{for } x \le 1;
\\
\displaystyle
-\frac{1}{4}\left[\log \frac{1+\sqrt{1-1/x}}{1-\sqrt{1-1/x}}
-i\pi \right]^2
&,& \textrm{for } x > 1.
\end{array}
\right.
\end{equation}
%-----------------
We note that there is no $W$-loop contribution to $h_\Phi \to \gamma\gamma$
since $h_\phi$ does not couple with the $W$ boson at the tree level.
However the charged Higgs boson contributes to the two-photon decay width
via the charged Higgs loop with the $h_\Phi H^+ H^-$ coupling
%-----------------
\begin{equation}
\lambda_{h_\Phi H^+ H^-} = 
\tilde{\lambda}_1 \sin^2 \beta
+\tilde{\lambda}_2 \cos^2 \beta
+\sqrt{2} \mu_\Phi \sin\beta \cos\beta/v_\Phi,
\label{lamhphm}%
\end{equation}
%-----------------
normalized by $v_\Phi$.   The charged Higgs contribution to $h_\Phi \rightarrow 
\gamma\gamma$  gets smaller when the charged Higgs boson becomes heavier, 
showing the typical decoupling behavior. 

We find that in a reasonable parameter set, the charged Higgs contribution is
not so large. However, its contribution could become more important 
if the extra fermions get larger masses. In this work, we do not consider
the charged Higgs contribution, which is more model-dependent.

Numerically, the decay width of $h_\Phi \to gg$ could be ${\cal O}(10)$ GeV
for large Yukawa coupling $Y \approx 5-10$ and small $m_f$ (exotic fermion mass).
But for $y=1$, it is less than about $0.5$ GeV. 
On the other hand, the decay width of $h_\Phi \to \gamma\gamma$
is at most ${\cal O}(0.1)$ GeV even for large Yukawa coupling and small $m_f$.
For $y=1$, the decay width is less than $5\times 10^{-3}$ GeV in the entire
region of $m_f$. Without extra decay channels of $h_\Phi$, 
we cannot achieve $O(10)$ GeV 
decay width for $h_\Phi$ for large $m_f ~( \gtrsim 500$ GeV).

The decay rate for $h_\Phi \to Z \gamma$ is given by
%-----------------
\begin{equation}
\Gamma[h_\Phi \to Z \gamma] =
\frac{\alpha m_W^2 m_{h_\Phi}^3}{128 \pi^2 v_H^2 v_\Phi^2}
\left( 1- \frac{m_Z^2}{m_{h_\Phi}^2}\right)^3
\left|
\sum_{f^\prime}^{} N_{f^\prime} \frac{Q_{f^\prime} v_{f^\prime}}{c_w}
A_{1/2}^H (\tau_{f^\prime}, \lambda_{f^\prime}) 
\right|^2,
\end{equation}
%-----------------
where $N_{q^\prime}=N_c$, $N_{l^\prime}=1$, and
$v_f=2 I_f^3 - 4 Q_f x_w$.
Here we do not consider the contribution of the charged Higgs boson like 
the $h_\Phi \to \gamma\gamma$ decay.

The function $A_{1/2}^H(\tau,\lambda)$ is defined by
%-----------------
\begin{equation}
A_{1/2}^H(\tau,\lambda) = I_1(\tau^{-1},\lambda^{-1})
-I_2(\tau^{-1},\lambda^{-1}),
\end{equation}
%-----------------
where
%-----------------
\begin{eqnarray}
I_1(x,y) &=& \frac{xy}{2(x-y)} + \frac{x^2 y^2}{2(x-y)^2}
\left[ f(x^{-1})-f(y^{-1})\right]
+ \frac{x^2 y}{(x-y)^2}
\left[ g(x^{-1})-g(y^{-1})\right],
\\
I_2(x,y)&=& -\frac{xy}{2(x-y)}
\left[f(x^{-1})-f(y^{-1})\right]
\end{eqnarray}
%-----------------
with 
%-----------------
\begin{equation}
g(x) = \left\{ 
\begin{array}{ccc}
\sqrt{x^{-1}-1} \arcsin \sqrt{x} &,& \textrm{for } x \le 1 \, ,
\\
\displaystyle
\frac{\sqrt{1-x^{-1}}}{2}
\left[\log \frac{1+\sqrt{1-1/x}}{1-\sqrt{1-1/x}}
-i\pi \right]^2
&,& \textrm{for } x > 1 \, ,
\end{array}
\right.
\end{equation}
%-----------------
and $f(x)$ is defined in eq.~(\ref{lamhphm}).

\begin{figure}[!t]
%\begin{center}
{\epsfig{figure=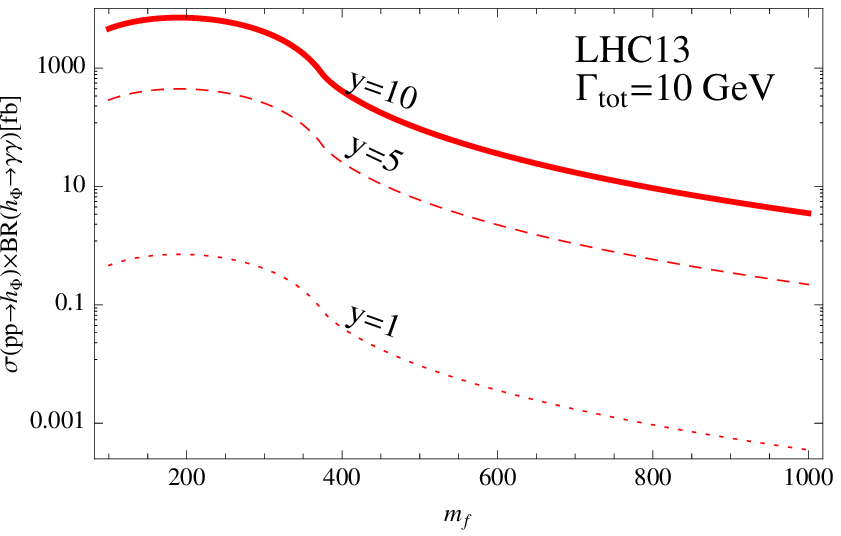,width=0.5\textwidth}}
{\epsfig{figure=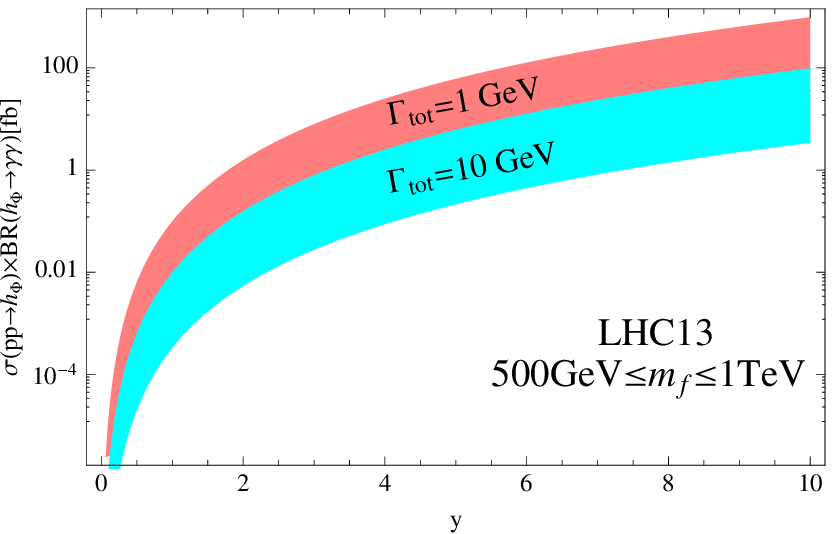,width=0.5\textwidth}}
\vspace{-0.5cm}
\caption{
(Left): $m_f$ vs. the diphoton signal via the gluon-gluon fusion at the LHC@13TeV. 
 The total decay widths is fixed at $\Gamma_{\rm tot}=10$ GeV.
 (Right): $y$ vs. the diphoton signal for different values of the $\Gamma_{\rm tot}$ and 
 for $500$ GeV $\leq m_f \leq$ 1 TeV. }
\label{fig1}
%\end{center}
\end{figure}

\begin{figure}[!t]
{\epsfig{figure=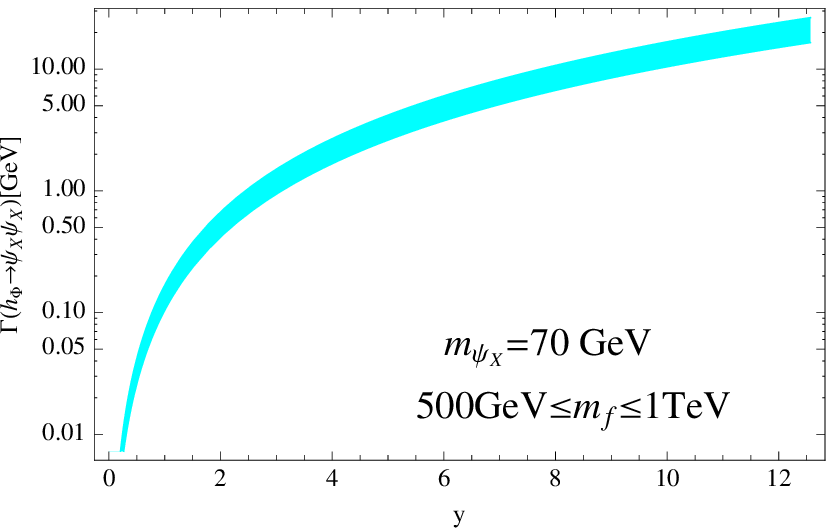,width=0.5\textwidth}}{\epsfig{figure=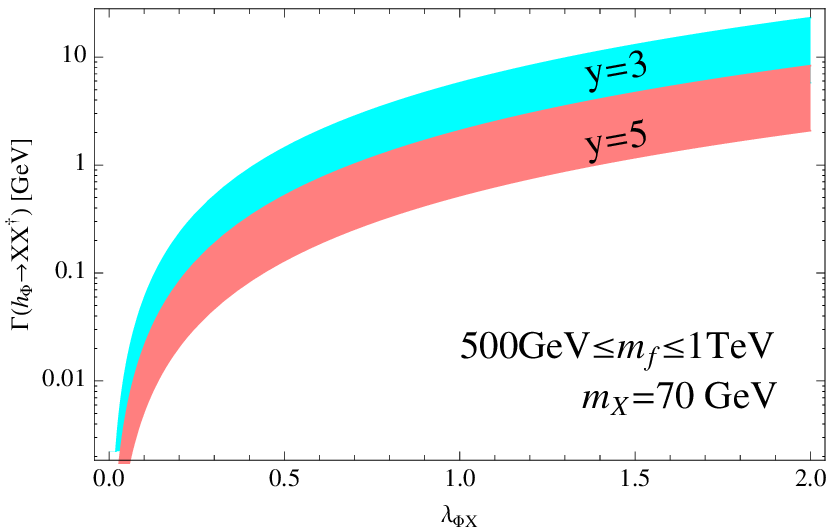,width=0.5\textwidth}}
\vspace{-0.5cm}
\caption{
$y $ vs. invisible decay width of $h_\Phi$ (GeV) in the fermionic DM scenario (left) and scalar DM scenario (right).The vector-like fermion mass is between $500$ GeV and $1$ TeV on the cyan 
and pink bands. The dark matter masses are $70$ GeV in the both cases. }
\label{fig2}
\end{figure}

Besides, $h_{\Phi}$ can decay to the extra particles, as well as 
the dark matter particles.
Now, let us simply define the extra decay width $\Delta \Gamma$ (GeV) and the total decay 
width ($\Gamma_{\rm tot}$) of $h_{\Phi}$  could be given by
\beq
\Gamma_{\rm tot}=\Delta \Gamma+\Gamma[h_\Phi \to gg] +\Gamma[h_\Phi \to \gamma \gamma]+\Gamma[h_\Phi \to \gamma Z].   
\eeq
The diphoton excess requires ${\cal O}(10)$-GeV $\Gamma_{\rm tot}$ and  ${\cal O}(10)$ fb diphoton signal 
at $\sqrt s =13$ TeV. This means that large $\Delta \Gamma$ is necessary to reproduce the excess.
In figure~\ref{fig1}, we see the required Yukawa coupling $(y=\sqrt 2 m_{f}/v_{\Phi})$, where $m_f$ is the mass of the extra fermions, and the diphoton signal at LHC13.  In the left panel, the total decay width is fixed at $10$ GeV, which can be readily achieved by allowing
the invisible decay of $h_\Phi$ into a pair of DM particles (see figure~\ref{fig2} and the related discussions).
In the right, $\Gamma_{\rm tot}=1$ GeV (pink) and $\Gamma_{\rm tot}=10$ GeV (cyan) are shown, when $m_f$ is between $500$ GeV and $1$ TeV.
Note that we need large Yukawa coupling $y \approx 5-10$ for $m_f > 400$ GeV 
 in order to get the correct 
size of the production cross section  for $pp \rightarrow h_\Phi \rightarrow \gamma\gamma$. 
Even if the total decay width is $1$ GeV, we still need large Yukawa coupling, as we see in the right panel.

%%------------------------------------------------------------------------------
%------------------------------------------------------------------------------

Now, we discuss the extra decay width $\Delta \Gamma$ in greater detail.   
First of all, $h_\Phi$ can decay into $Z_H Z_H$, a pair of leptophobic U(1)$^{'}$ gauge bosons.
If there exists the mixing between $\hat{Z}$ and $\hat{Z}_H$ bosons,  
it can also decay into $Z Z$ and $Z Z_H$ through the mixing at the tree level.
But the decay widths will be suppressed by small gauge coupling $g_H \lesssim O(0.1)$  and the small 
$Z-Z_H$ mixing, which are required from the search for the extra neutral gauge boson~\cite{Ko}.
There is also contribution to $h_\Phi \to W W, Z Z, Z Z_H, Z_H Z_H$  from the loop diagrams of extra 
fermions  which have $SU(2)_L$ and/or $U(1)^\prime$ quantum numbers.
The last two channels are suppressed by the small $g_H$ coupling, while  the decay widths in the first two 
channels may be the same order as $\Gamma[h_\Phi \to \gamma\gamma]$. 
Numerically, their contribution  to the total decay width would be negligible like 
$\Gamma[h_\Phi \to \gamma\gamma]$ in the case of $\Gamma_\textrm{tot} \sim O(10)$ GeV.

%First of all, $h_\Phi$ can decay into $Z Z , Z Z_H , Z_H Z_H $,
%if there exists the mixing between $\hat{Z}$ and $\hat{Z}_H$ bosons.  
%But it will be suppressed by small gauge coupling $g_H \lesssim O(0.1)$ 
%and the small $Z-Z_H$ mixing.
%Therefore we will ignore $h_\Phi \rightarrow Z Z, Z Z_H , Z_H Z_H$. 

Next, we consider the $h_\Phi$ decay into two scalar bosons ($s_i s_j$),
$s_i = h, H, H^\pm, A$,  and into two Majorana fermions ($\psi_X \psi_X$), where $\psi_X$  
is one of dark matter candidates, if kinematically allowed.  These new channels are unique aspects   
of  our model  due to the 2HDM nature of leptophobic U(1)$'$ models. We can enhance the decay width
of $h_\Phi$ not by ad hoc way but by underlying gauge theory which is anomaly-free.  
The decay widths for $h_\Phi \to s_i s_j$ and $h_\Phi \to \psi_X \psi_X$  are given by
%-----------------
\begin{eqnarray}
\Gamma[h_\Phi \to s_i s_i] &=& 
\frac{\lambda_{h_\Phi s_i s_i}^2 }{32 \pi } \frac{v_\Phi^2}{m_{h_\Phi}} 
\sqrt{1-4 x_i}, \label{eq;ss}
\\
\Gamma[h_\Phi \to s_i s_j] &=& 
\frac{\lambda_{h_\Phi s_i s_j}^2 }{16 \pi} \frac{v_\Phi^2}{m_{h_\Phi}} 
\lambda^{1/2}(1, x_i, x_j)  ({\rm for}~i\neq j) \ ,   \label{eq;ss-1}  \\
\Gamma[h_\Phi \to \psi_X \psi_X] &=& \frac{y_X^2m_{h_\Phi}}{8 \pi} \{1 - 4(m^2_{\psi_{X}}/m_{h_\Phi}^2)\}^{\frac{3}{2}},
\end{eqnarray}
%-----------------
where $x_i = m_{s_i}^2/m_{h_\Phi}^2$ and $\lambda(a,b,c)=a^2+b^2+c^2-2ab-2bc-2ca$. $y_X$ and $m_{\psi_{X}}$ are the Yukawa coupling with $h_{\Phi}$ and 
the mass of $\psi_X$. 

The non-zero entries of $\lambda_{h_\Phi} s_i s_j$ are given by 
%-----------------
\begin{eqnarray}
\lambda_{h_\Phi h h} &=&
\tilde{\lambda}_1 \sin^2 \alpha_h
+ \tilde{\lambda}_2  \cos^2 \alpha_h
+ \frac{\mu_\Phi}{\sqrt{2} v_\Phi} \sin \alpha_h \cos \alpha_h,
\label{eq;hh}
\\
\lambda_{h_\Phi H H} &=&
\tilde{\lambda}_1 \cos^2 \alpha_h
+ \tilde{\lambda}_2 \sin^2 \alpha_h
- \frac{\mu_\Phi}{\sqrt{2} v_\Phi} \sin \alpha_h \cos \alpha_h,
\label{eq;HH}
\\
\lambda_{h_\Phi h H} &=&
-\tilde{\lambda}_1  \sin 2\alpha_h
+ \tilde{\lambda}_2  \cos 2\alpha_h
- \frac{\mu_\Phi}{\sqrt{2}v_\Phi} \cos 2\alpha_h,
\label{eq;hH}
\\
\lambda_{h_\Phi A A} &=&
\frac{v_\Phi^2}{v_\Phi^2+(v_H \sin\beta \cos \beta)^2}
\nonumber \\
&\times&
\left(
\tilde{\lambda}_1  \sin^2 \beta
+\tilde{\lambda}_2  \cos^2 \beta
+\frac{\mu_\Phi}{\sqrt{2}v_\Phi} \sin\beta \cos\beta
+2 \lambda_\Phi v_H \sin\beta \cos\beta
\right),
\label{eq;AA}
\end{eqnarray}
%-----------------
where $A$ is the pseudoscalar boson.  The coupling $\lambda_{h_\Phi H^+ H^-}$ has been
defined in eq.~(23) in the context of $h_\Phi \rightarrow \gamma \gamma$.

In addition, we find a dark matter candidate among the natural fermions \cite{Ko}.
Assuming that the Yukawa couplings are flavor-independent, we can explicitly calculate the 
DM mass and the Yukawa coupling with $h_\Phi$.
figure~\ref{fig2} shows the partial decay widths of $h_\Phi$ to two dark matter particles in the fermionic DM scenario (left) and scalar DM scenario (right).
As we see, the $h_\Phi$ invisible decay cannot be so large in the fermionic DM case, although the branching ratio is relatively larger than  the diphoton decay width. 
In the region where the perturbativity holds, the invisible decay width of $h_\Phi$ is at most 
about 10 GeV. In this case, if the total decay width of the 750 GeV excess is 
confirmed to be about 45 GeV, then other decay channels like 
$h_\Phi \to h H , H H , A A $ must be comparable to or dominant over the invisible decay 
\footnote{The charged  scalar mass is constrained by $B\rightarrow X_s \gamma$ in type-II Higgs doublet 
model,  and should satisfy $m_{H^\pm} \gtrsim 500$ GeV \cite{Misiak:2015xwa}.  
Therefore $h_\Phi \rightarrow H^+ H^-$ is kinematically forbidden in our model. } 
(see the end of this section for more discussion on this point).
On the other hand, the invisible decay in the scalar DM case 
can be dominant, if $\lambda_{\Phi X}$ is ${\cal O}(1)$.

In our model, there is a massive gauge boson ($Z_H$), which dominantly gets the mass from 
the nonzero VEV of $\Phi$. The extra fermions also get the mass from the VEV, so
there is a relation between the $Z_H$ mass and $m_f$.
Approximately, it can be evaluated as
\beq
M_{Z_H} \approx \frac{g_H}{y} m_f.
\eeq
The diphoton excess suggests ${\cal O} (1)$ $y$, and then $M_{Z_H}$ is at least ${\cal O}(100)$ GeV,  
because $g_H$ could not be ${\cal O}(1)$ to evade the stringent bound from the dijet signal.
Another strong constraint is from $\rho$ parameter, as discussed in ref. \cite{Ko}.
In this scenario, $Z_H$ is light, so the coupling should be small.
As discussed in refs. \cite{Ko-0,Ko,Hisano}, the $\rho$ parameter is deviated from $1$
at the tree-level, because of $Z$-$Z_H$ mixing. The bound is roughly estimated as \cite{Hisano}
\beq
\frac{g_H}{g_Z} \frac{M^2_Z}{|M^2_{Z_H}-M^2_Z|} \lesssim 0.004.
\eeq
Then ${\cal O} (100)$-GeV $M^2_{Z_H}$ requires $g_H \lesssim {\cal O} (0.1) \times g_Z$,
which may be too small to enhance the branching ratio of $h_{\Phi} \to Z_H Z_H$.

There are some experimental constraints relevant to our scenario.
Since the $h_\Phi$ is produced from the gluon fusion copiously 
at the LHC, the dijet production can severely constrain our scenario.
The bound on the dijet production at LHC@8TeV is about 2 pb at CMS~\cite{CMS-dijet}.
By imposing 
%-----------------
\begin{equation}
\sigma(gg\to h_\Phi \to gg) \lesssim 2~\textrm{pb},
\end{equation}
%-----------------
we find that the mass of the exotic quarks should be larger than
400 GeV for the Yukawa coupling $y=5$ and 600 GeV for $y=10$.

Next, we consider the diboson channels. First, the $h_\Phi$ can decay into
$hh$. Then this channel is constrained by the experimental data at LHC@8TeV~\cite{Aad:2015uka}:
%-----------------
\begin{equation}
\sigma (gg  \to h_\Phi \to hh) \lesssim 10~{\rm fb}.
\label{hhbound}%
\end{equation}
%-----------------
The decay width of $h_\Phi \to hh$ strongly depends on the model parameters. 
For example, for $\alpha_h=0$ and $\mu_\Phi \sim v_H$,
the branching ratio of $h_\Phi \to hh$ is ${\cal O}(0.1)$ for $\tan\beta\sim 1$,
while it could be ${\cal O}(0.01)$ for $\tan\beta \sim 10$.
Actually, the bound (\ref{hhbound}) requires $Br(h_\Phi\to hh) \lesssim 0.01$.

We note that there are other diboson channels, $h_\Phi \to ZZ, WW$.
The bounds from LHC8 for the $WW$ production~\cite{Aad:2015kna} 
and $ZZ$ production~\cite{Aad:2015agg} are
%-----------------
\begin{eqnarray}
\sigma(gg\to h_\Phi \to WW) &\lesssim& 40~\textrm{fb}, \\
\sigma(gg\to h_\Phi \to ZZ) &\lesssim& 10~\textrm{fb}, 
\end{eqnarray}
%-----------------
respectively. In our model, $h_\Phi$ does not interact with $W$ and $Z$ bosons
at the tree-level directly so that the decay channels are suppressed by loop diagrams
or $Z$-$Z_H$ mixing, which must be small. Since the loop diagrams of the extra
leptons which has the SM $SU(2)_L$ quantum number are
dominant, the decay width for $h_\Phi \to WW$ and $ZZ$ would be the same order
as $\gamma\gamma$ or less. Therefore, the bound for the diboson channels $WW$ and $ZZ$ 
would be acceptable.

Finally, we consider the $Z\gamma$ production channel. At LHC8, the bound
on the production is~\cite{Aad:2014fha}
%-----------------
\begin{eqnarray}
\sigma(gg\to h_\Phi \to Z\gamma) &\lesssim& 3.8~\textrm{fb}.
\end{eqnarray}
%-----------------
From the numerical analysis, we find that 
the extra fermion mass should be larger than about 200 (400) GeV for 
$y=5 \, (10)$, respectively. For $y=1$, the cross section is less than 1 fb.
They are weaker than the bound from the dijet production.

If the invisible decay of $h_\Phi$ is dominant, the monojet search at the LHC
would give most stringent constraints on the models which may explain
the diphoton excess. The NLO correction to $gg \rightarrow h_\Phi$ is very involved,
and beyond the scope of this paper. Here we try to make a qualitative argument on the 
monojet + $\ET$ constraints.  The monojet + $\ET$ signal will be generated by the parton level
processes: (i) the initial state radiation of gluon in $gg\to h_\Phi$ and 
$qg \to qgg$ followed by $gg\to h_\Phi$ via triangle diagrams
and (ii) $gg\to h_\Phi g$ via box diagrams.
The type (i) will mainly generate a monojet in the beam direction with low $p_T$
and may be removed by the $\ET$ cut. They are also suppressed by an extra 
$g_s$. The type (ii) could generate high $p_T$ monojet and should be constrained
by the data on monojet + $\ET$.   If we use a naive dimensional analysis, its rate would be 
suppressed by $\sim \alpha_s / (4 \pi)$ compared with the rate for $g g\rightarrow h_\Phi$ 
from a triangle diagram. 
At the LHC@8TeV, $\sigma ( gg \rightarrow h_\Phi ) \approx 2$ pb, so that we would expect
that $\sigma ( gg \rightarrow h_\Phi g ) \approx 0.2$ pb, which satisfies the bound $\lesssim 0.8$ pb 
derived in ref.~\cite{Franceschini:2015kwy}.  
For more definitive conclusion on this issue, we have to perform more detailed analysis. 

The dijet + $\ET$ process may constrain our model. For the vector boson fusion
process, this dijet + $\ET$ occurs via a parton level process,
\begin{eqnarray*}
q \bar{q^{'}} & \rightarrow & q \bar{q^{'}} + Z_H Z_H , \ \ \ {\rm followed ~by}
 \\
Z_H Z_H & \rightarrow & h_\Phi \rightarrow X X^\dagger,  
\end{eqnarray*}
where $X$ is a dark matter particle. Considering the current bounds
on the $m_{Z_H}$ and $g_H$, this $Z_H Z_H$ fusion production could be 
neglected safely.

Another dijet + $\ET$ events could arise from $g g$ fusion to $h_\Phi$ beyond the leading 
order, 
\begin{eqnarray*}
g g & \rightarrow & g g + g g (\rightarrow h_\Phi ) \\
qq ({\rm or}~ q \bar{q} ) & \rightarrow & q q ({\rm or}~q \bar{q} ) + g g (\rightarrow h_\Phi ) \\ 
q g ({\rm or}~\bar{q} g) & \rightarrow & q g ({\rm or}~\bar{q} g) + g g (\rightarrow h_\Phi )
\end{eqnarray*}
which will be $O(\alpha_s)$ suppressed compared with what we have studied in the 
earlier part of this section, namely $gg \rightarrow h_\Phi$.  And the dijets in these processes
will be mainly in the beam directions with low $p_T$ and we expect that they will be removed 
by  the $p_T$ cuts.

Finally, let us comment on other decay modes of $h_\Phi$ to two extra scalars, such 
as $h_\Phi \to H h$, $HH$, and $AA$, whose decay rates eqs. (\ref{eq;ss}) and (\ref{eq;ss-1}) 
are determined by the dimensionless couplings in the Higgs potential, eqs.~(\ref{eq;hh})--(\ref{eq;AA}).  
In principle, these extra decay could be sizable up to $\Delta \Gamma \approx 
{\cal O}(10)$ GeV. 
$H$ and $A$ mainly decay into the $b \overline{b}$ state, and so the final states in the 
$h_\Phi \rightarrow H h$ decay channel would be $bb \overline{b}  \overline{b}$. 
This channel is not strongly constrained by the present data, and 
is one of the promising signals of our scenario.

%%------------------------------------------------------------------------------
%------------------------------------------------------------------------------

%%%%%%%%%%%%%%%%%%%%%%%%%%%%%%%%%%%%%%%%%%%%%%%
\section{Summary}
\label{sec:summary}
%%%%%%%%%%%%%%%%%%%%%%%%%%%%%%%%%%%%%%%%
 
In this paper we interpret the recently reported diphoton excess at 750 GeV
 in terms of a new singlet scalar boson $h_\Phi$ that originates from spontaneous breaking of 
leptophobic U(1)$^{'}$ embedded in E$_6$ grand unification.  
A {\bf 27}-dimensional fundamental representation of E6 gauge group contains one family 
of SM chiral fermions, as well as 11 more chiral fermions, some of which are vectorlike under
the SM $SU(2)_L \times U(1)_Y$ gauge symmetry.  Anomaly cancellation is automatic 
in this model, and exotic fermions are chiral under U(1)$^{'}$ so that their masses arise 
entirely from spontaneous breaking of U(1)$^{'}$ symmetry by the nonzero VEV of $\Phi$.
The observed diphoton excesses are attributed to  
$gg \rightarrow h_\Phi \rightarrow \gamma\gamma$.  
The vectorlike exotic fermions are chiral under new U(1)$^{'}$ gauge symmetry and 
their masses are generated only by spontaneous gauge symmetry. Therefore their loop 
effects would be protected from the decoupling theorem, like the top quark loop contributions
to $h\rightarrow gg, \gamma\gamma$, etc..  In our model, $h_\Phi$ can decay into a pair 
of DM, as well as two scalar bosons such as $h h, H h , A A $, etc.. In particular the $H h$ 
final state can have $O(10)$ GeV decay width, making one of the dominant decay channels 
of  $h_\Phi$. 

If the diphoton excess at 750 GeV with large decay width $\sim$ 45 GeV  is confirmed in the next LHC
run,  our model predicts that there should be new vectorlike quarks and leptons around $\sim O(1)$TeV
(or lighter for vectorlike leptons), whose collider signatures would be similar to the squark/slepton searches
within the $R$-parity conservation, namely dijet + $\ET$ or dilepton + $\ET$.  
Also additional scalar bosons will be present too, and one of the main decay channels 
of $h_\Phi$   would be $H h$ final state.    In our model the production
cross sections for exotic fermions will be larger than the sfermion productions because they are spin-1/2
fermions.  
In addition, there will be a new leptophobic (baryonic) gauge boson $Z_H$
whose mass could be still as low as a few GeV for a few hundred GeV $v_\Phi$
and $g_H \lesssim O(0.1)$, which is mainly constrained by the decay width
of the $Z$ boson, $\Gamma_Z$ and the $\rho$ parameter~\cite{Ko}.
DM will be either spin-1/2 fermion or spin-0 scalar,
and they will be baryonic in a sense that they have interactions with the nuclei through $Z_H$ exchanges.
DM phenomenology within this model in the context of 750 GeV diphoton excess will be presented elsewhere.

%---------------------------------------------------------------------------
%%%%%%%%%%%%%%%%%%%%%%%%%%%%%%%%%%%%
\section*{Acknowledgments}
%%%%%%%%%%%%%%%%%%%%%%%%%%%%%%%%%%%%
%---------------------------------------------------------------------------
This work is supported in part by National Research Foundation of Korea (NRF) 
Research Grant NRF-2015R1A2A1A05001869, and by the NRF grant 
funded by the Korea government (MSIP) (No. 2009-0083526) through Korea 
Neutrino Research Center at Seoul National University (P.K.). 
The work of Y.O. is supported by Grant-in-Aid for Scientific research
from the Ministry of Education, Science, Sports, and Culture (MEXT),
Japan, No. 23104011. 
The work of C.Y. was supported by the Ministry of Science and Technology (MoST) of Taiwan
under grant number 101-2112-M-001-005-MY3.

%---------------------------------------------------------------------------

\appendix

%\vspace{-1ex}

%%%%%%%%%%%%%%%% References    %%%%%%%%%%%%%%%%%%%%

\end{document}